# Charge Transfer Excited State Contributions to Polarity Dependent Ferromagnetism in ZnO Diluted Magnetic Semiconductors

Kevin R. Kittilstved, William K. Liu, and Daniel R. Gamelin*

*Department of Chemistry, University of Washington, Seattle, Washington 98195-1700, U.S.A.*

**e-mail: Gamelin@chem.washington.edu*

**Abstract**. Future spintronics technologies based on diluted magnetic semiconductors (DMSs) will rely heavily on a sound understanding of the microscopic origins of ferromagnetism in such materials. Recent discoveries of room-temperature ferromagnetism in wide-bandgap DMSs hold great promise, but this ferromagnetism remains poorly understood. Here we demonstrate a close link between the charge transfer electronic structures and polar high-$T_C$ ferromagnetism of $TM^{2+}$:ZnO DMSs ($TM^{2+}$ = 3d ions). Trends in ferromagnetism across the 3d series of $TM^{2+}$:ZnO DMSs predicted from their charge transfer energies reproduce experimental trends well. These results provide a unified basis for understanding both n- and p-type ferromagnetic oxide DMSs.

## Introduction

Diluted magnetic semiconductors (DMSs) are attracting intense interest for potential new device applications in spin-based information processing technologies. For practical spintronics applications, ferromagnetic DMSs with Curie temperatures ($T_C$) greatly exceeding room temperature will be required. Theoretical predictions[1,2] of high-$T_C$ ferromagnetism in DMSs of GaN and ZnO stimulated efforts to develop these and related wide-bandgap materials.[3] In many regards, ZnO is the archetypical wide band gap semiconductor. Compared to other oxides that have been investigated as DMSs (e.g. $TiO_2$, $SnO_2$), ZnO is relatively well behaved from an experimental point of view, affording the possibility for extensive systematic experimentation and data analysis. Well-defined doping and defect chemistries, suitability for transparent high-power high-temperature applications, and ability to lase or emit spontaneously at UV wavelengths combine to make ZnO attractive for many potential device applications.[4] For spintronics applications, the relatively long room-temperature spin coherence time of n-type ZnO is advantageous.[5] Additionally, the potential to generate both p- and n-type ZnO of low resistivity makes bipolar spintronics based on ZnO a realistic possibility, and recent reports of both hole-[6-9] and electron-mediated[10-12] ferromagnetism in ZnO DMSs are encouraging in this regard. The microscopic origins of this high-$T_C$ ferromagnetism remain poorly understood, however. A detailed understanding of high-$T_C$ ferromagnetism in wide-bandgap DMSs is required in order to harness functionality for device applications. The development of such an understanding has emerged as among the most important challenges in modern magnetism.[13,14]

In this article, we demonstrate that ferromagnetism in ZnO DMSs is intimately related to the charge transfer (CT) electronic structures of the magnetic impurity ions. Electronic absorption, magnetic circular dichroism (MCD), and photocurrent action spectroscopies are used to identify and assign CT excited states and to analyze the properties of these electronic configurations related to defect-mediated ferromagnetism. Of particular interest are the opposite polarities predicted[2] and observed[6] for ferromagnetic $Co^{2+}$:ZnO (n-type) and $Mn^{2+}$:ZnO (p-type). From experimentally calibrated CT energies and configuration interaction considerations, trends in ferromagnetism across the 3d series of n- and p-type $TM^{2+}$:ZnO DMSs are predicted and compared to experiment. The predicted trends reproduce experimental trends remarkably well, confirming the role of CT configurations in the ferromagnetism of $TM^{2+}$:ZnO DMSs and, by extension, wide-bandgap DMSs as a class.

## Experimental Trends in $TM^{2+}$:ZnO DMS Ferromagnetism

Several groups have reported ferromagnetism in $TM^{2+}$:ZnO DMSs.[3] To minimize potential sources of experimental error, some groups have compared different $TM^{2+}$:ZnO DMSs prepared under identical conditions. Ueda et al. examined a series of n-type $TM^{2+}$:ZnO ($TM^{2+}$ = $Co^{2+}$, $Mn^{2+}$, $Cr^{2+}$, and $Ni^{2+}$) films prepared by pulsed laser deposition (PLD).[15] High-$T_C$ ferromagnetism was observed only in $Co^{2+}$-doped films and not in any of the others. Venkatesan et al. prepared a broader series of n-type ZnO DMSs by PLD and observed two maxima in the 300K ferromagnetic saturation moments, one at $Co^{2+}$:ZnO ($d^7$) and the other at $V^{2+}$:ZnO ($d^3$).[12,14] $Mn^{2+}$:ZnO ($d^5$) showed little or no ferromagnetism under the sample preparation conditions of either laboratory. Kittilstved et al. observed the same trend in natively n-type $Co^{2+}$:ZnO and $Mn^{2+}$:ZnO prepared using an oxidative direct chemical route that unambiguously precluded formation of metal precipitates.[6] The chemical approach also allowed investigation of the influence of nitrogen, a p-type dopant in ZnO. When nitrogen was introduced during materials preparation, $Mn^{2+}$:ZnO showed strong ferromagnetism whereas $Co^{2+}$:ZnO showed none.[6,9,16] When plotted in table format as in Fig. 1, the opposite carrier polarities of ferromagnetic $Co^{2+}$:ZnO and $Mn^{2+}$:ZnO become strikingly apparent. This polarity dependent ferromagnetism, referred to here as polar ferromagnetism, is shown below to derive from specific electronic structural properties of the $Co^{2+}$ and $Mn^{2+}$ ions in the ZnO host semiconductor.

The experimental trends described above and in Fig. 1 are in general agreement with most theoretical predictions, despite the use of a variety of different theoretical models.[1,2,14,17,18] A key property common to all of the models describing ZnO DMS ferromagnetism is strong electronic coupling between the magnetic ions and charge carriers at the Fermi level. Since isovalent $TM^{2+}$ doping of ZnO does not itself introduce carriers, carriers in $TM^{2+}$:ZnO DMSs are associated with additional shallow donor or acceptor defects. The dopant-carrier exchange energy is parameterized variously as $J_{sd}$, $J_{pd}$, $N_0\alpha$, $N_0\beta$, etc.,[1,14,19] depending on the specific interaction and model under consideration. Microscopically, these exchange coupling parameters account for partial carrier delocalization onto the magnetic dopant. In a configuration interaction picture, this implies mixing between the relevant localized and charge separated electronic configurations of the magnetic impurity ion. The striking differences between $Co^{2+}$:ZnO and $Mn^{2+}$:ZnO in Fig. 1 indicate that $TM^{2+}$ electronic structures must regulate the abilities of holes and electrons to delocalize onto these dopants. Specifically, the data in Fig. 1 suggest that the resonances described by equations 1a,b are relevant, whereas those of equations 1c,d are not.

$$Co^{2+} + e^-_{donor} \leftrightarrow Co^+ \quad (1a)$$

$$Mn^{2+} + h^+_{acceptor} \leftrightarrow Mn^{3+} \quad (1b)$$

$$Co^{2+} + h^+_{acceptor} \not\leftrightarrow Co^{3+} \quad (1c)$$

$$Mn^{2+} + e^-_{donor} \not\leftrightarrow Mn^+ \quad (1d)$$

The forward direction in equation 1a describes formal transfer of a shallow donor's electron ($e^-_{donor}$) onto $Co^{2+}$ to form $Co^+$, whereas in equation 1b it describes transfer of a shallow acceptor's hole ($h^+_{acceptor}$) onto $Mn^{2+}$ to form $Mn^{3+}$. The reverse directions describe formal CT processes involving the reduced ($Co^+$) or oxidized ($Mn^{3+}$) magnetic dopants. As equilibria, equations 1a,b thus describe dopant-donor/acceptor hybridization. This hybridization is the pivotal feature determining $T_C$ in theoretical models describing DMS ferromagnetism. In the spin-split donor impurity band model for n-type ferromagnetic DMSs,[14] a phenomenological dopant effective radius scaling parameter $(r_c^{eff}/r_o)^3$ was introduced to account for hybridization enhancement of the s-d exchange parameter ($J_{sd}$) required to increase the predicted $T_C$ from 1 to ~500K. The actual extent of hybridization was small, and 1-2% electron transfer to the dopant was considered sufficient,[14] i.e., the ground state lies far to the left in equation 1a, a conclusion consistent with ferromagnetic X-ray MCD data showing multiplet structure characteristic of tetrahedral $Co^{2+}$ in ZnO.[20] Similarly, in the Zener model description of p-type ferromagnetic DMSs, the high $T_C$ predicted for p-type $Mn^{2+}$:ZnO derived from its large p-d exchange parameter ($N_0\beta$),[1] the magnitude of which is determined primarily by p-d hybridization.[19] *Ab initio* LSDA-DFT calculations[2,17,18] have predicted ferromagnetism in p-type $Mn^{2+}$:ZnO that results from partial delocalization of $h^+_{acceptor}$ onto $Mn^{2+}$, which imparts manganese 3d character at the Fermi level, but $e^-_{donor}$ did not delocalize onto $Mn^{2+}$ and consequently no ferromagnetism was predicted for n-type $Mn^{2+}$:ZnO.[2] To understand such charge delocalization processes, and to develop a comprehensive model that describes both n- and p-type high-$T_C$ ferromagnetic ZnO DMSs, it is imperative to understand the CT electronic structures of these DMSs.

## Charge transfer electronic structures

***(I) Spectroscopic results.*** The CT electronic structures of $Co^{2+}$:ZnO and $Mn^{2+}$:ZnO were investigated using optical spectroscopic probes. In general, two types of CT transitions are anticipated in ZnO DMSs. In one case, an electron may be promoted from the $TM^{2+}$ ion to ZnO-based acceptor orbitals of the conduction band (CB), and in the other case an electron may be promoted to the $TM^{2+}$ ion from ZnO-based donor orbitals of the valence band (VB). The state generated at the CT electronic origin is expected to involve a semiconductor electron (or hole) that is loosely bound to the charged impurity by Coulombic forces,[21,22] possibly with a large effective radius. Treating the semiconductor lattice as a ligand to the dopant ion, these transitions are formally metal-to-ligand conduction band CT ($ML_{CB}CT$) and ligand valence band-to-metal CT ($L_{VB}MCT$) transitions, as summarized in equation 2.

$$TM^{2+} \rightarrow TM^{+} + h^{+}_{VB} \qquad (L_{VB}MCT) \qquad (2a)$$

$$TM^{2+} \rightarrow TM^{3+} + e^{-}_{CB} \qquad (ML_{CB}CT) \qquad (2b)$$

Figure 2 shows electronic absorption spectra, photocurrent internal quantum efficiencies (IQEs),[23] and MCD spectra of paramagnetic $Co^{2+}$:ZnO and $Mn^{2+}$:ZnO. The intense absorption and MCD feature at ~28,000cm$^{-1}$ in both DMSs arises from the first excitonic configuration of ZnO at the band edge. The structured intensity of $Co^{2+}$:ZnO centered at 16,000cm$^{-1}$ is the spin-orbit split $^4A_2 \rightarrow {}^4T_1(P)$ ligand field band of $Co^{2+}$ doped substitutionally into wurtzite ZnO.[24,25]

Two additional sub-bandgap features are observed in the MCD and photocurrent action spectra of $Co^{2+}$:ZnO that were not evident by absorption, one a strong negative MCD peak at ~25,000cm$^{-1}$ and the other a very weak, broad band that extends down to 14,000cm$^{-1}$. Both features show S = 3/2 MCD saturation magnetization confirming that they come from isolated paramagnetic $Co^{2+}$ ions. The breadths of the two bands indicate they are due to CT transitions. Their origin from two distinct CT transitions is evident from their ~4-fold different photocurrent IQEs (Fig. 2b).[23] The higher energy CT band is assigned as the $L_{VB}MCT$ transition (equation 2a) on the basis of its relationship to a similar band in $Ni^{2+}$:ZnO.[25] The lower energy CT band is the $ML_{CB}CT$ transition (equation 2b).[23] These CT assignments are supported by calculations using Jørgensen's optical electronegativity model,[26,27] in which CT transition energies are related to differences in donor (D) and acceptor (A) optical electronegativities ($\chi_{opt}$(D) and $\chi_{opt}$(A)) after taking into account differences in multi-electron spin-pairing energies (SPEs) and dopant ligand field effects between the ground and excited states (equation 3).

$$E_{CT} = 30{,}000\text{cm}^{-1}(\chi_{opt}(\text{D})-\chi_{opt}(\text{A})) + \Delta SPE \pm 10Dq \qquad (3)$$

The SPEs account for Coulomb and electron-electron exchange interactions at the 3d ions and are evaluated using the Racah parameters B and C. From equation 3, $\chi_{opt}(Co^{2+}) = 1.9$, $\chi_{opt}(ZnO_{VB}) = 2.4$, $\chi_{opt}(ZnO_{CB}) = 1.1$, and the ligand-field parameters of $Co^{2+}$ in ZnO (B = 775cm$^{-1}$, C/B = 4.5, Dq =

390cm$^{-1}$),[23,25] the energies of the $L_{VB}MCT$ and $ML_{CB}CT$ transitions are estimated to be ~25,000 and 11,300cm$^{-1}$, respectively, in reasonable agreement with the experimental energies.

In the spectra of $Mn^{2+}$:ZnO (Fig. 2d-f), only one sub-bandgap CT band is observed, a photoactive state having an associated pseudo-A-term MCD signal centered at ~24,000cm$^{-1}$. From equation 3, the $L_{VB}MCT$ and $ML_{CB}CT$ transitions of $Mn^{2+}$:ZnO ($\chi_{opt}(Mn^{2+}) = 1.45$, B = 596cm$^{-1}$, C/B = 6.5, Dq = 420cm$^{-1}$) are predicted to occur at ~49,000 and 24,000cm$^{-1}$, respectively, leading to assignment of this band as the $ML_{CB}CT$ transition.[9] Its assignment to ligand field transitions is ruled out by its high molar extinction coefficent ($\varepsilon(Mn^{2+}) \approx 950M^{-1}cm^{-1}$ at 24,000cm$^{-1}$ and 300K, compared with $\varepsilon(Mn^{2+}) \approx 1 - 10M^{-1}cm^{-1}$ at ~24,850cm$^{-1}$ anticipated for the $^4T_1(G)$ ligand-field excited state[9]). The very high energy predicted for the $L_{VB}MCT$ transition in $Mn^{2+}$:ZnO agrees with that estimated from analysis of $Mn^{2+}$:ZnO X-ray absorption data (~52,500 ± 12,100cm$^{-1}$).[28]

Two key observations from Fig. 2 pertain to the magnetism of these ZnO DMSs: (i) both $Co^{2+}$:ZnO and $Mn^{2+}$:ZnO possess CT excited states immediately below the ZnO band edge; (ii) the identities of the CT excited states at the band edge are different for $Co^{2+}$:ZnO and $Mn^{2+}$:ZnO, the former being an $L_{VB}MCT$ state and the latter being an $ML_{CB}CT$ state. As described below, observation (i) relates to the existence of high-$T_C$ ferromagnetism mediated by shallow donors or acceptors, whereas observation (ii) relates to its polarity.

***(II) Wavefunctions.*** The CT spectroscopic data also provide information about wavefunctions. Excited states leading to detectable photocurrents in the $TM^{2+}$:ZnO photovoltaic cells generate both electrons and holes at the electrode surfaces. Although the absolute photocurrent IQEs depend on cell engineering parameters,[23] the ratio of photocurrent IQEs for the two CT transitions in $Co^{2+}$:ZnO ($IQE_{25000}/IQE_{20000} \approx 4$ from Fig. 2b) is due to the intrinsic branching ratios for charge separation in the respective CT excited states and reflects differences in mixing between localized (equation 4a,b) and delocalized (equation 4c) excited state configurations for the carrier transferred to the dopant upon excitation. The configuration interaction wavefunctions for the two CT excited states in the perturbation limit are given by equation 4d, where the mixing coefficient $c_{n,3} = H_{n3}/\Delta E_{n,3}$, and $H_{n3}$ and $\Delta E_{n,3}$ are the off-diagonal CT electronic coupling matrix element and the energy difference between the relevant CT and excitonic configurations, respectively.

$$\psi_1(L_{VB}MCT):\ TM^{+} + h^{+}_{VB} \tag{4a}$$

$$\psi_2(ML_{CB}CT):\ TM^{3+} + e^{-}_{CB} \tag{4b}$$

$$\psi_3(\text{Excitonic}):\ TM^{2+} + e^{-}_{CB} + h^{+}_{VB} \tag{4c}$$

$$\psi_n' = \psi_n + c_{n,3}\psi_3 \quad (4d)$$

The relatively small energetic difference ($\Delta E_{1,3} \approx 2600\ cm^{-1}$, Fig. 2) between the $L_{VB}$MCT and excitonic states in $Co^{2+}$:ZnO imparts partial excitonic character to the $L_{VB}$MCT state and reflects a relatively small binding energy for the photo-generated bound electron, favoring carrier escape. Conversely, the substantially greater energetic difference ($\Delta E_{2,3} \approx 13,400\ cm^{-1}$, Fig. 2) between the $ML_{CB}$CT and excitonic states does not allow extensive mixing of the two, so the photo-generated hole is strongly bound to the cobalt and charge recombination is favored. The ratio of localization energies for the last bound charge carriers in the two CT states in $Co^{2+}$:ZnO ($\Delta E_{2,3}/\Delta E_{1,3} \approx 5$) is similar to the ratio of photocurrent IQEs measured for these two excited states (~4, Fig. 2b), and both CT IQEs are smaller than that for excitonic excitation in pure ZnO, consistent with attribution of CT photocurrent to admixture of $\psi_3$ into the CT wavefunctions $\psi_1$ and $\psi_2$, as described by equation 4d. Quantitative analysis of the CT oscillator strengths yields estimated mixing coefficients $c_{1,3} = 0.17 \pm 0.01$ and $c_{2,3} = 0.043 \pm 0.003$ for $Co^{2+}$:ZnO,[23] and $c_{2,3} = 0.19 \pm 0.03$ for $Mn^{2+}$:ZnO.

***(III) Born cycle analysis.*** The CT energies of dopants in unstable oxidation states can be analyzed using Born thermodynamic cycles. Following McClure et al.,[29] the energy of a $ML_{CB}$CT transition for $TM^{2+}$ in ZnO as defined by equation 2b is given by equation 5, where $I_3$(TM) is the 3rd ionization potential of the dopant, $\Delta V(\text{site}) = V_3 - V_2$ is the difference in total potential for $TM^{3+}$ and $TM^{2+}$ at the $Zn^{2+}$ crystal site, and $\chi$ is the electron affinity of the semiconductor.

$$E_{M^{2+}LCT} = I_3(\text{TM}) - \Delta V(\text{site}) - \chi \quad (5)$$

Similarly, the $L_{VB}$MCT energy involving the one-electron oxidized dopant ($TM^{3+}$) is given by equation 6, where $E_g$ is the band gap energy of the host crystal.

$$E_{LM^{3+}CT} = E_g + \Delta V(\text{site}) - I_3(\text{TM}) + \chi \quad (6)$$

Concatenation of these two processes gives $TM^{2+} \rightarrow TM^{3+} + e^-_{CB} \rightarrow TM^{2+} + e^-_{CB} + h^+_{VB}$, which is identical to excitonic excitation (equation 4c), and the sum of equations 5 and 6 yields $E_{M^{2+}LCT} + E_{LM^{3+}CT} = E_g$. Corresponding relationships hold for the reverse order of promotions, $TM^{2+} \rightarrow TM^{+} + h^+_{VB} \rightarrow TM^{2+} + h^+_{VB} + e^-_{CB}$, for which $E_{LM^{2+}CT} + E_{M^{+}LCT} = E_g$. $E_{LM^{3+}CT}$ and $E_{M^{+}LCT}$ are thus identical to $\Delta E_{2,3}$ and $\Delta E_{1,3}$ from equation 4a-c, respectively. This model assumes that $e^-_{CB}$ and $h^+_{VB}$ are far from their sources in the CT excited states, a condition that is never actually fulfilled. Although additional corrections should be considered, these corrections are small relative to the photon energies under consideration and empirically the model accounts for CT energies reasonably well.[29] From these identities and the CT spectroscopic data in Fig. 2, $E_{M^{+}LCT}$ and $E_{LM^{3+}CT}$ can therefore be deduced for cobalt and manganese in ZnO. From this analysis, the transition $Co^{+} \rightarrow Co^{2+} + e^{-}_{CB}$ occurs at very low energy ($\Delta E_{1,3} \approx 2600 cm^{-1}$) as does the transition $Mn^{3+} \rightarrow Mn^{2+} + h^{+}_{VB}$ ($\Delta E_{2,3} \approx 3400 cm^{-1}$), whereas those involving $Co^{3+}$ ($\Delta E_{2,3} \approx 13,400 cm^{-1}$) and $Mn^{+}$ ($\Delta E_{1,3} \approx -21,600 cm^{-1}$) do not. This analysis is summarized schematically in Fig. 3 and numerically in Table 1.

**Relationship between CT Electronic Structure and Ferromagnetism**

***(I) Dopant-donor/acceptor hybridization.*** In this section, we assume the basic framework of the recent spin-split donor impurity band model[14] and focus on the key microscopic property governing defect-mediated ferromagnetism in this model, namely hybridization between the magnetic dopant and the defect so that carriers at the Fermi level are partially delocalized onto the magnetic dopant (equation 1, and accounted for in ref. 14 using the scaling parameter $(r_c^{eff}/r_o)^3$). Whereas the model in ref. 14 was developed exclusively for donor defects, the kernel of the model also applies to acceptor defects as shown below. From perturbation theory, dopant-defect hybridization is parameterized by the mixing coefficient, $c_{DD}$ (equation 7). For effective hybridization, the energy difference between dopant and defect levels ($\Delta E_{DD}$) must be small and the resonance integral ($H_{DD}$) must be large.

$$c_{DD} = H_{DD}/\Delta E_{DD} \qquad (7)$$

To analyze $\Delta E_{DD}$, the thermodynamics of the four dopant-defect charge delocalization processes described by equation 1a-d are plotted in Fig. 4 for commonly invoked shallow donors and acceptors in ZnO (Shallow donor: interstitial zinc ($Zn_i$).[30] Shallow acceptor: substitutional nitrogen ($N_O^{2-}$).[4] See Table 1.). These plots expose cases where charge delocalization is approximately thermoneutral (i.e. small $\Delta E_{DD}$). The major conclusion drawn from Fig. 4 is that dopant-defect electron transfer approaches thermoneutrality only when the reduced (or oxidized) dopant approaches a potential similar to that of the shallow donor (or acceptor). From Fig. 4 and Table 1, near-thermoneutral electron transfer with shallow donor or acceptor defects can occur in p-type $Mn^{2+}$:ZnO ($|\Delta E_{DD}| \approx 0.22 eV$) and n-type $Co^{2+}$:ZnO ($|\Delta E_{DD}| \approx 0.27 eV$), but not in n-type $Mn^{2+}$:ZnO ($|\Delta E_{DD}| \approx 2.7 eV$) or p-type $Co^{2+}$:ZnO ($|\Delta E_{DD}| \approx 1.7 eV$). The small $\Delta E_{DD}$ values for p-type $Mn^{2+}$:ZnO and n-type $Co^{2+}$:ZnO favor effective hybridization (equation 7) and hence high-$T_C$ ferromagnetism, whereas the large $\Delta E_{DD}$ values for n-type $Mn^{2+}$:ZnO and p-type $Co^{2+}$:ZnO limit hybridization and are not favorable for ferromagnetism. The polarity conditions for ferromagnetism predicted by this CT analysis are thus in agreement with the experimental results summarized in Fig. 1.

The resonance integral of equation 7 is proportional to the spatial overlap of dopant and defect wavefunctions[31,32] which, for fixed concentrations of dopants and defects across the transition metal series, in turn relates to the effective Bohr radii of the limiting charge carriers on each center. Equation 8 allows effective Bohr radii for the $TM^{n+}$ and donor/acceptor species in Fig. 4 to be estimated.[33]

$$r_B = \frac{\hbar}{\sqrt{2m^* E_b}} \tag{8}$$

In equation 8, m* is the effective mass of the relevant carrier (in ZnO, $m^*_e \approx 0.24m_e$ and $m^*_h \approx 0.45m_e$) and $E_b$ is the binding energy of the last bound carrier. Table 1 summarizes the relevant binding energies and Bohr radii. From the CT analysis (Fig. 3), $Co^+$ is a shallow donor and $Mn^{3+}$ a shallow acceptor in ZnO, with carrier binding energies $E_b \approx \Delta E_{n,3}$, whereas $Co^{3+}$ and $Mn^+$ are both energetically misaligned with the band structure of ZnO. The Bohr radii for $Mn^+$ and $Co^{3+}$ are both substantially smaller than those of $Co^+$, $Zn_i$, $Mn^{3+}$, or $N^{2-}$. Experimental confirmation of this comparison is obtained from the photocurrent IQEs of Fig. 2b, which show that $e^-$ hopping is considerably more favorable than $h^+$ hopping in photoexcited $Co^{2+}$:ZnO. Dopant-donor/acceptor hybridization is thus favored by energetic proximity of both to the appropriate band edge of the host semiconductor. In general, although deeper donors or acceptors may give thermoneutral electron transfer with deeper magnetic dopants, their smaller Bohr radii would require correspondingly increased concentrations to achieve sufficient overlap, at which point short-range antiferromagnetic superexchange interactions become problematic.

***(II) Trends across the $TM^{2+}$:ZnO DMS series.*** To test the broader applicability of this CT analysis for understanding the magnetic properties of ZnO DMSs, trends in $c_{DD}$ were investigated. For ZnO DMSs prepared under identical conditions, the major changes in $c_{DD}$ across the series of 3d $TM^{2+}$ ions occur in the energy denominator, $\Delta E_{DD}$. CT energies for the 3d $TM^{2+}$ series were calculated from Pauling electronegativities and literature ligand field parameters using equation 3 (see Supplementary Information). Equation 3 predicts substantial negative $ML_{CB}CT$ energies for $Sc^{2+}$ and $Ti^{2+}$ in ZnO, consistent with instability of these ions to spontaneous oxidation. $Ti^{2+}$ and $Sc^{2+}$ are therefore not considered in subsequent analysis.

Figure 5 compares the results of the CT analysis with available experimental and theoretical magnetic data across the same series. To retain generality, $|1/\Delta E_{n,3}|$ is plotted rather than $|1/\Delta E_{DD}|$, but both yield identical trends for fixed shallow defects. The trends predicted from the CT analysis are in excellent overall agreement with the experimental data,[12,15] particularly considering the crudeness of the CT energy calculations from equation 3 and the potential experimental uncertainties. The CT predictions are also in excellent agreement with available LSDA-DFT results across the series.[2] This agreement is a very strong confirmation that partial dopant-donor/acceptor charge delocalization (i.e. exchange mediated by shallow bound carriers) is responsible for ferromagnetism in ZnO DMSs. Of the three factors from equation 3 (electronegativities, spin pairing energies, and ligand field splittings), spin pairing energies are primarily responsible for the positions of the maxima and minima in Fig. 5. A comprehensive model of both hole- and electron-mediated ferromagnetism in ZnO DMSs therefore must account for Coulomb and electron-electron exchange interactions of the 3d ions.

Although carriers are implicated, this analysis does not imply macroscopic conductivity. The magnetic phase diagrams for carrier-mediated ferromagnetism in ZnO DMSs may possess a region in which the material is both insulating and ferromagnetic. Such a region has been proposed previously for n-type oxide DMSs[14] and has been observed experimentally in manganese-doped GaAs,[34] where the critical dependence of ferromagnetism on carriers is generally accepted.[1]

**Conclusion**

To summarize, high-$T_C$ ferromagnetism in ZnO DMSs is closely linked to the CT electronic structures of the transition metal dopants. The following conditions have been identified as generally conducive to dopant-donor/acceptor hybridization at the Fermi level, and hence to high-$T_C$ ferromagnetism:

(i) approximately thermoneutral dopant-defect electron transfer (i.e., small $\Delta E_{DD}$)

(ii) energetic proximity of the reduced or oxidized dopant to the semiconductor band structure (i.e., small $\Delta E_{n,3}$)

Satisfaction of these conditions can be recognized experimentally by the appearance of CT excited states in close proximity to the band edge in electronic absorption, MCD, or photocurrent action spectra of the DMS, or predicted by calculations of CT energies that take into account electron-electron repulsion at the 3d transition metal ions. Moreover, assignment of the CT excitation ($L_{VB}MCT$ versus $ML_{CB}CT$) can reveal the polarity of the carrier mediating ferromagnetism (n- versus p-type, respectively). Although this study has focused on ZnO DMSs, similar electronic structural properties likely also govern ferromagnetism in other wide-bandgap DMSs. Indeed, CT absorption and MCD intensities very near the band edge are observed in $Co^{2+}$:$TiO_2$,[35] $Ni^{2+}$:$SnO_2$,[36] and $Cr^{3+}$:$TiO_2$,[37] all three of which show high-$T_C$ ferromagnetism under appropriate conditions. Careful analysis of the CT excited states in these and other wide-bandgap DMSs, coupled with continued development of sophisticated theoretical models, can therefore be anticipated to provide valuable new insights into the electronic structural origins of polar high-$T_C$ ferromagnetism in this important class of materials.

**Methods.** Nanocrystalline films of $Mn^{2+}$:ZnO (up to 2% $Mn^{2+}$) and $Co^{2+}$:ZnO (up to 3.5% $Co^{2+}$) for magnetic[6] and photovoltaic[23] measurements were prepared as described previously.[6,23] Dopant concentrations were determined by inductively coupled plasma atomic emission spectrometry (ICP-AES, Jarrel Ash model 995). Electronic absorption spectra were collected using a Cary 5E spectrophotometer (Varian). MCD spectra were collected using a UV/VIS/NIR spectrophotometer constructed from an Aviv 40DS with a sample compartment modified to house a high-field superconducting magneto-optical cryostat mounted in the Faraday configuration. MCD intensities were measured as the absorbance difference $\Delta A = A_L - A_R$ (where L and R refer to left and right circularly polarized photons) and are reported as $\theta(\deg) = 32.9\Delta A/A$. Magnetic susceptibility measurements were performed using a Quantum Design MPMS-5S SQUID magnetometer. The net diamagnetic background was subtracted from the raw magnetization data (see Supplementary Information). Photocurrent action spectra were measured using a $\mu$Autolab Type II potentiostat (Eco Chemie B.V.) integrated with the Aviv 40DS spectropolarimeter and converted to IQE as described previously.[23] Details of the calculations of charge transfer energies for Fig. 5 are provided as Supplementary Information.

**Acknowledgments.** Financial support from the NSF (PECASE DMR-0239325 and ECS-0224138), the Research Corporation (Cottrell Scholar), the Dreyfus Foundation (Teacher/Scholar), and the NSF-IGERT program at U.W. (to W.K.L.) is gratefully acknowledged.

**Supplementary Information.** describing the details of the calculations for Fig. 5 accompanies the paper and is available online (web address to be added)

**Table 1.** Donor or acceptor properties for reduced or oxidized transition metal dopants in ZnO, estimated from equation 8 and the data in Fig. 2. The properties of commonly invoked shallow donors and acceptors are included for comparison.

| Donor or Acceptor | Carrier type | $E_b$ ($cm^{-1}$)[a] | $E_b$ (eV)[a] | $r_B$ (nm) | ref. |
|---|---|---|---|---|---|
| $Mn^{3+}$:ZnO | $h^+$ | 3400 | 0.42 | 0.45 | b |
| $Mn^{+}$:ZnO | $e^-$ | -21,600 | -2.68 | 0.24 (but unstable) | b |
| $Co^{3+}$:ZnO | $h^+$ | 13,400 | 1.66 | 0.23 | b |
| $Co^{+}$:ZnO | $e^-$ | 2600 | 0.32 | 0.70 | b |
| $N_O^{2-}$:ZnO | $h^+$ | 1600 | 0.20 | 0.65 | 4 |
| $Zn_i^0$:ZnO | $e^-$ | 240 | 0.03 | 2.31 | 30 |

(a) From $\Delta E_{n,3}$ for TM dopants. (b) This work.

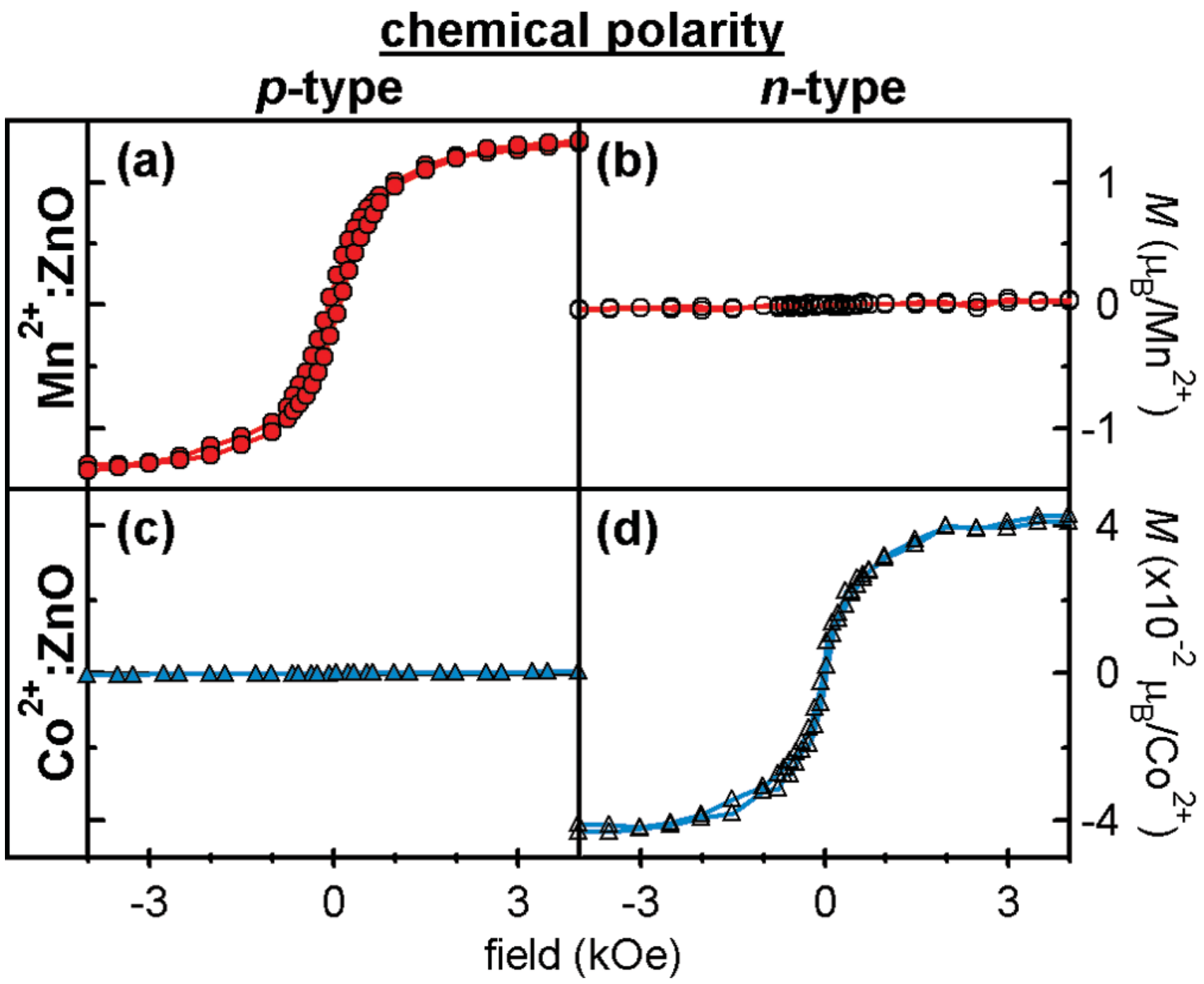

**Figure 1.** 300 K magnetization data for 0.2% $Mn^{2+}$:ZnO and 3.5% $Co^{2+}$:ZnO films prepared by direct chemical synthesis with or without addition of nitrogen. **(a)** 0.2% $Mn^{2+}$:ZnO with added nitrogen, **(b)** 0.2% $Mn^{2+}$:ZnO without added nitrogen, **(c)** 3.5% $Co^{2+}$:ZnO with added nitrogen, and **(d)** 3.5% $Co^{2+}$:ZnO without added nitrogen.

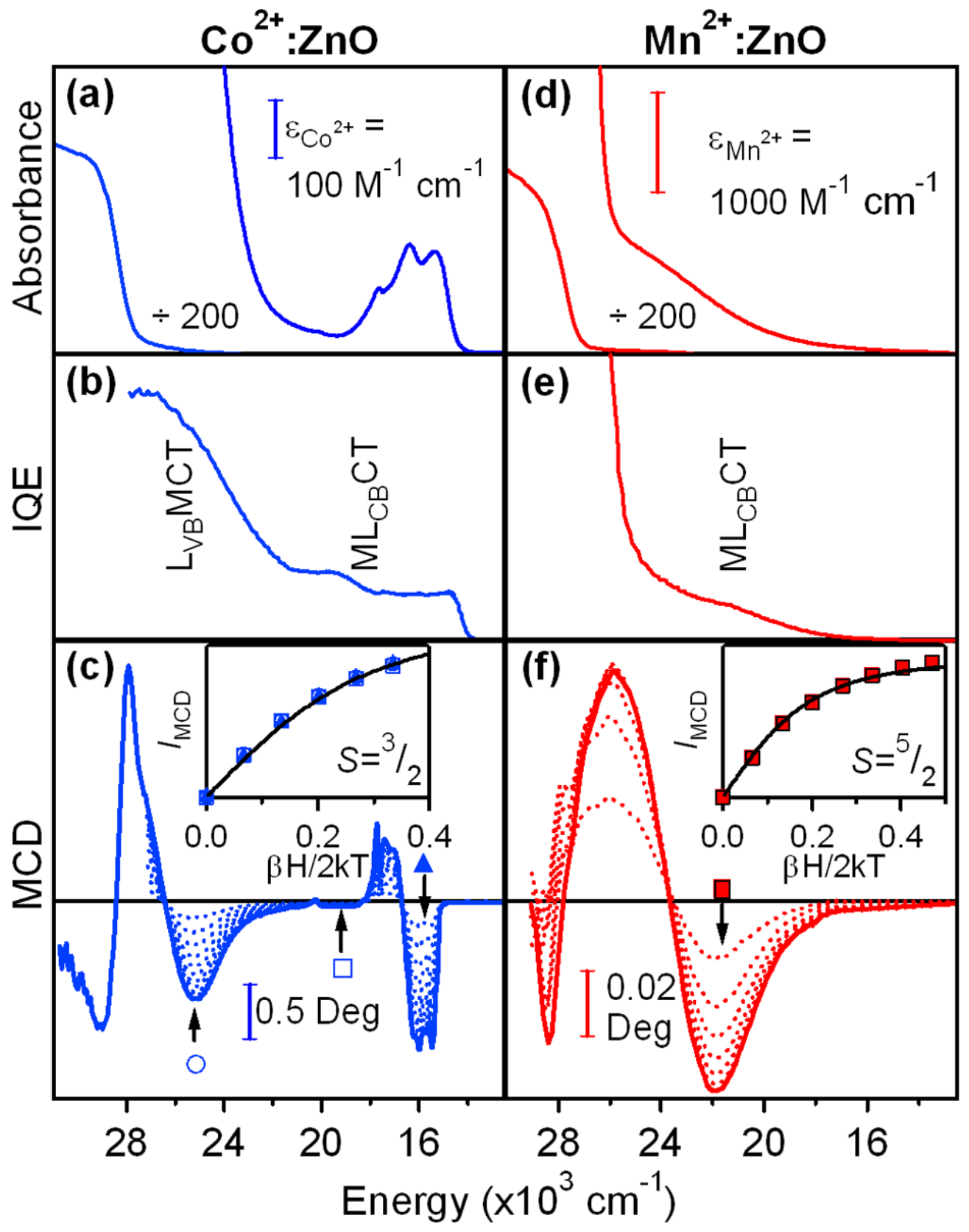

**Figure 2.** Electronic absorption, photocurrent internal quantum efficiency, and magnetic circular dichroism spectroscopic data for $Co^{2+}$:ZnO and $Mn^{2+}$:ZnO. **(a,d)** 300K electronic absorption spectra. **(b,e)** Photocurrent internal quantum efficiencies of photovoltaic cells. **(c,f)** Variable-field 5K MCD spectra. The insets show MCD saturation magnetization data measured at the energies marked, superimposed on $S = {}^3/_2$ and $S = {}^5/_2$ Brillouin curves.

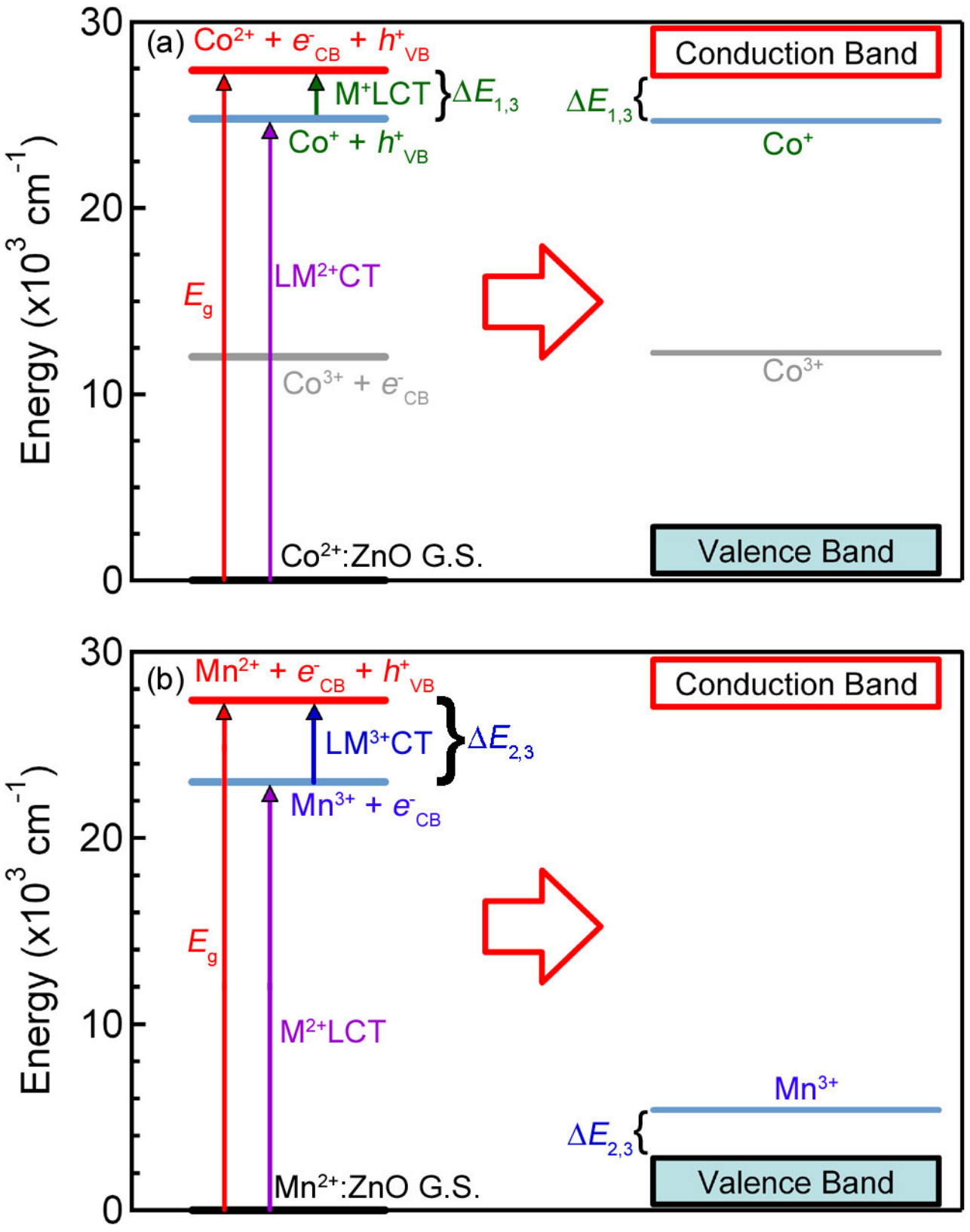

**Figure 3.** Schematic summary of the CT analysis for $Co^{2+}$- and $Mn^{2+}$-doped ZnO DMSs showing the relationship between excited state energies and donor or acceptor energies derived from the Born cycle analysis. Top: $Co^{2+}$:ZnO. Bottom: $Mn^{2+}$:ZnO. From this analysis, $Co^+$ should be a shallow donor and $Mn^{3+}$ a shallow acceptor in ZnO.

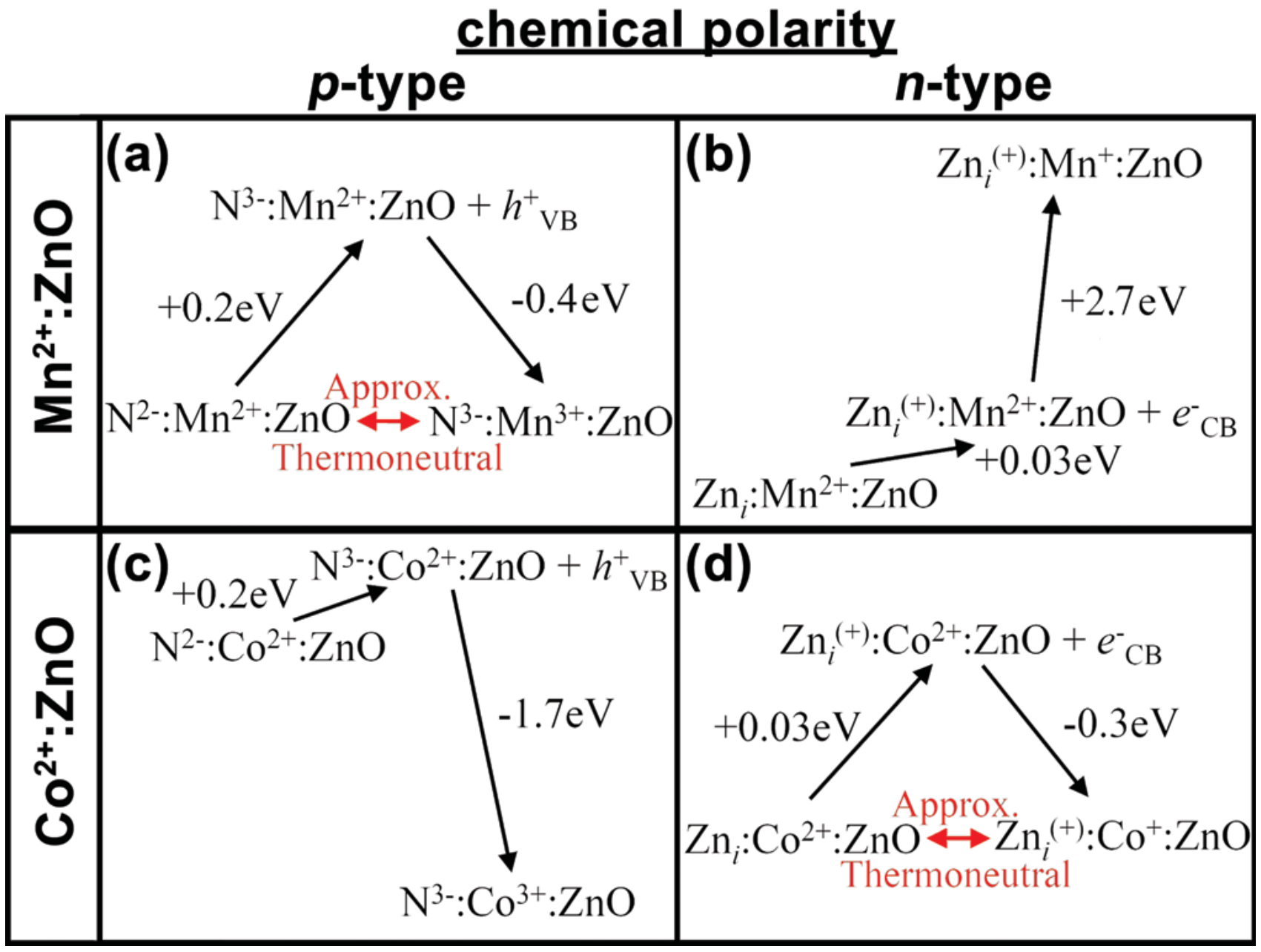

**Figure 4.** Thermodynamics of dopant-donor/acceptor charge transfer processes determined from charge transfer analysis and the binding energies of common shallow donors ($Zn_i$) or acceptors ($N_O^{2-}$) in ZnO. The resulting values of $|\Delta E_{DD}|$ are: (a) 0.22, (b) 2.7, (c) 1.7, and (d) 0.27 eV. Electron transfer is approximately thermoneutral only for p-type $Mn^{2+}$:ZnO and n-type $Co^{2+}$:ZnO.

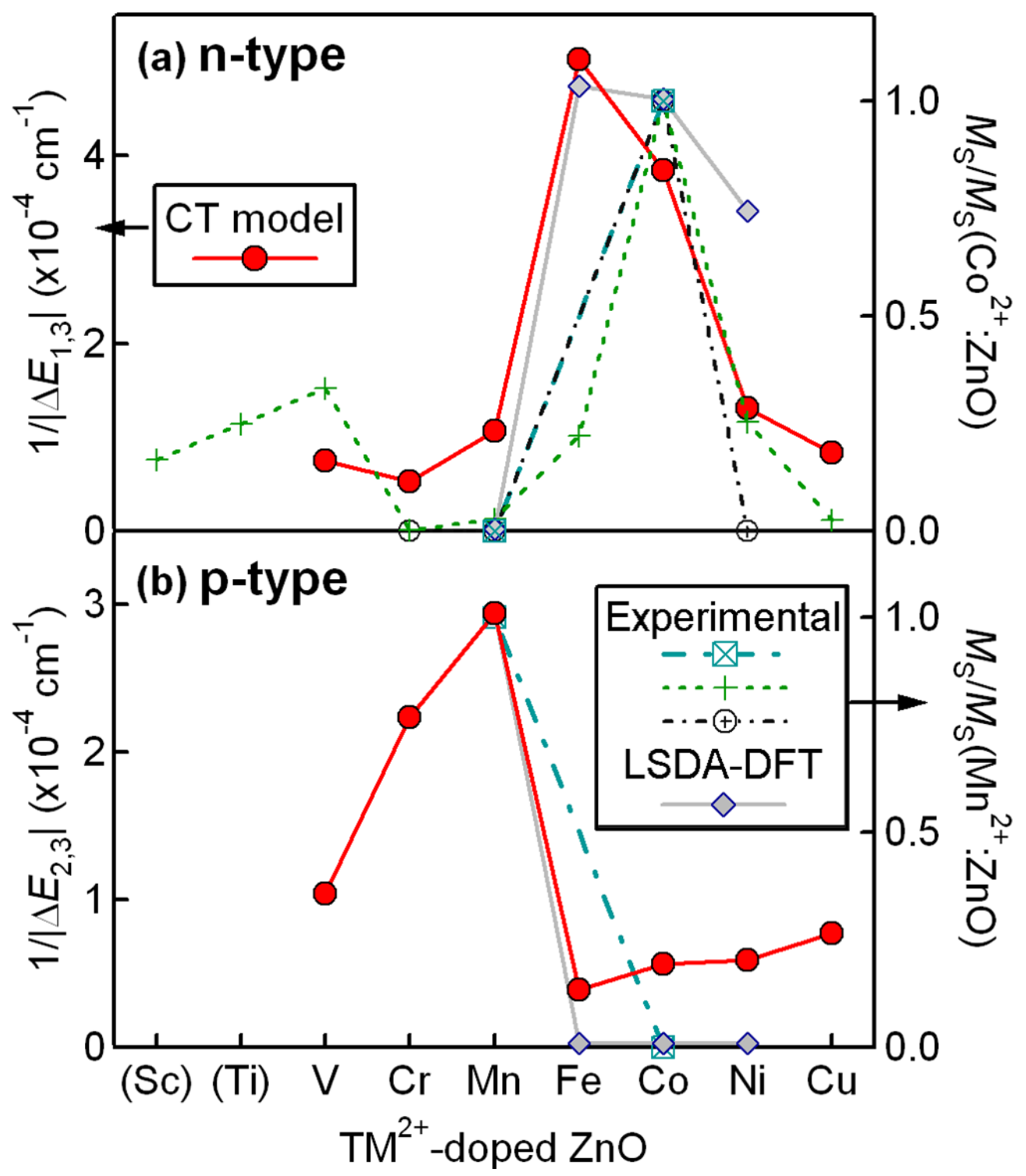

**Figure 5.** Results of charge transfer analysis for the series of $TM^{2+}$:ZnO DMSs. Calculated trend (●) for **(a)** n-type polarity and **(b)** p-type polarity. Charge transfer energies were calculated using equation 3 and are plotted as $|1/\Delta E_{n,3}|$ (left abscissa). Experimental data are from refs. 15 (⊕), 12 (+), and from Fig. 1 (⊠). *Ab initio* LSDA-DFT results are from ref. 2 (◆). For comparison purposes, each literature data set is plotted normalized to the values for $Co^{2+}$:ZnO in (a) and $Mn^{2+}$:ZnO in (b) that were reported for the same data set (right abscissa). Sc and Ti are anticipated to be in the 3+ oxidation state and were therefore not included in the CT analysis.